\documentclass[aps,longbibliography,superscriptaddress,onecolumn,preprint,tightenlines]{revtex4-1}
\usepackage{graphicx}
\usepackage{epstopdf}
\usepackage{amssymb,amsmath}
\usepackage{relsize}
\usepackage{placeins}
\usepackage{mathtools}
\usepackage{setspace}
\usepackage{enumitem}
\usepackage{lineno}
\usepackage{tikz}
\usepackage{pgffor}
\usepackage{booktabs}   
\usepackage{lineno,hyperref}
\usepackage{comment}
\usepackage{nicefrac}
\usepackage{todonotes}

\def\xv{\boldsymbol{x}}

\def\uv{\boldsymbol{u}}

\def\tauv{\boldsymbol{\tau}}

\def\gammav{\boldsymbol{\gamma}}

\def\onedot{$\mathsurround0pt\ldotp$}
\def\cddot{
  \mathbin{\vcenter{\baselineskip.67ex
    \hbox{\onedot}\hbox{\onedot}}%
  }}%
\def\cdddot#1{
  \mathbin{\vcenter{\baselineskip.67ex
    \hbox{\onedot}\hbox{\onedot}\hbox{\onedot}%
  }}%
}

\def\onedot{$\mathsurround0pt\ldotp$}
  \def\cddot{
  \mathbin{\vcenter{\baselineskip.67ex
  \hbox{\onedot}\hbox{\onedot}}%
  }}%

\newlist{steps}{enumerate}{1}
\setlist[steps]{label=\textit{Step~\arabic*~}}

\graphicspath{{figures/}}

\begin{document}

\title{Direct numerical simulation of particle sedimentation in a Bingham fluid}

\author{A.R.~Koblitz}
\affiliation{Department of Physics, Cavendish Laboratory, J J Thomson Avenue, Cambridge CB3 0HE, UK}
\thanks{A.R.~Koblitz acknowledges financial support from the EPSRC Centre for Doctoral Training in Computational
Methods for Materials Science under grand EP/L015552/1.}
\thanks{This work was supported by Schlumberger Cambridge Research Limited.}
\email{ark44@cam.ac.uk}

\author{S.~Lovett}
\affiliation{Schlumberger Cambridge Research Limited, High Cross, Madingley Road, Cambridge CB3 0EL, UK}

\author{N.~Nikiforakis}
\affiliation{Department of Physics, Cavendish Laboratory, J J Thomson Avenue, Cambridge, CB3 0HE, UK}

\date{\today}

\begin{abstract}

    The settling efficiency, and stability with respect to settling, of a dilute suspension of infinite circular cylinders
    in a quiescent viscoplastic fluid is examined by means of direct numerical simulations with varying
    solid volume fraction, $\phi$, and yield number, $Y$. For $Y$ sufficiently large we find higher settling
    efficiency for increasing $\phi$, similar to what is found in shear-thinning fluids and opposite to what is
    found in Newtonian fluids. The critical yield number at which the suspension is held stationary in the
    carrier fluid is found to increase monotonically with $\phi$, while the transition 
    to settling is found to be diffuse: in the same suspension, particle clusters may 
    settle while more isolated particles remain arrested. In this regime, complex flow features are observed 
    in the sedimenting suspension, including the mobilization of lone particles by nearby sedimentation 
    clusters. Understanding this regime, and the transition to a fully arrested state, is relevant to many
    industrial and natural problems involving the sedimentation of viscoplastic suspensions under quiescent flow
    conditions.
\end{abstract}

\maketitle

\section{Introduction}
\label{sec:introduction}

The dispersion of coarse particles in complex (shear dependent) fluids is an important aspect of both natural and
industrial flows, for example, pyroclastic gravity currents \citep{Dufek2016}, proppant transport in hydraulic
fracturing \citep{Osiptsov2017}, and fresh cement slurries. Often the suspended phase is (negatively) buoyant and
its stability with respect to sedimentation during transport, or after flow cessation \citep{Santos2018}, is of
fundamental interest. Viscoplastic fluids, by virtue of a material yield stress, may support such a coarse
particle phase indefinitely. This is governed by a balance between the stress exerted on the fluid by the particle
and the fluid yield stress \citep{Beris1985}, described by the non-dimensional yield number $Y =
{\hat{\tau}_y}/{(\hat{\rho}_p - \hat{\rho}_f)\hat{\boldsymbol{g}}\hat{l}'}$,
where $\hat{\boldsymbol{g}}$ is gravitational acceleration, $\hat{\rho}_p$ is the particle density, $\hat{\rho}_f$
the fluid density, $\hat{l}'$ a characteristic length scale, and $\hat{\tau}_y$ the material yield stress. 
Two aspects of
buoyant particle transport in viscoplastic fluids stand out: the conditions for suspension stability, and the
sedimentation behaviour. Both of these have been well characterized for single particles but not yet for
suspensions, particularly under quiescent flow conditions.

With regards to stability, it has been shown that for an isolated particle there exists a critical yield 
number $Y=Y^\ast_0$ at which the
buoyancy force exerted by the particle is balanced by the fluid yield stress.
This has been the subject of many numerical and theoretical 
works \citep{Tokpavi2008,Beris1985,Jossic2001} and while there is
great variability amongst experimental studies \citep{Emady2013,Chhabra2007}, 
the theoretical $Y^\ast_0$ for a spherical particle has been
corroborated by \citet{Tabuteau2007}. Given the non-linear rheology of the suspending fluid a key question is
whether $Y^\ast_0$ is applicable to suspensions. There has been some work, predominantly numerical and
experimental, investigating model systems of two or more particles posed in the resistance 
sense---where flow is driven by a prescribed
velocity---rather than the more applicable mobility sense---where flow is driven by applied force---due to the 
intrinsic numerical and practical difficulties of the latter. It was found that particles near 
each other, particularly in the inline
configuration, experience a decreased drag force, from which it may be inferred that the same
configuration would exhibit a higher critical yield number than an isolated particle
\citep{Tokpavi2009,Liu2003,Jie2006}. The
theoretical work of \citet{Frigaard2017a} investigated this critical yield number for suspensions more directly
through the out-of-plane flow of uniformly distributed particle suspensions with prescribed uniform suspension
velocity. They inferred a volume fraction, $\phi$, dependent critical yield number, $Y^\ast_\phi$. 
However, the resistance
formulation has clear drawbacks in that individual particle velocities are prescribed \textit{a priori}. Recently,
\citet{Chaparian2018} investigated inline particle configurations for up to 5 particles in the mobility sense,
finding not only that their stability criterion is strongly influenced by separation distance but that particle
chains are unlikely to be stable sedimentation configurations.

Settling of suspensions in viscoplastic fluids can be categorized as static settling or dynamic settling,
depending on the background flow conditions (quiescent in the former). It is well known in the oil industry that
background shear enhances settling \citep{Childs2016} in shear-thinning fluids. For viscoplastic fluids,
\citet{Merkak2009} and \citet{Ovarlez2012} demonstrated shear-induced settling in fluids regardless of the yield
number, showing that particles settle as soon as the fluid yield stress is overcome by macroscopic shear. The 
latter advocated a suspension settling function, in conjunction with a Newtonian hindering function, incorporating
an effective viscosity based on the (applied) macroscopic shear rate. This framework has recently been adopted in
a model for solids dispersion in hydraulic fracturing flows \citep{Hormozi2017}.

In quiescent background conditions, i.e.~in the absence of applied macroscopic shear, a different approach is
needed. Single particles settling in viscoplastic fluids under quiescent conditions have been investigated
extensively in viscous and inertial regimes, both experimentally and numerically
\citep{Atapattu1990,Yu2007,Wilson2003,Arabi2016}. Empirical terminal velocity models have been developed
\citep{Wilson2003,Arabi2016} and it has been suggested that these may be combined with Newtonian hindering
functions in dispersion models such as \citet{Kaushal2013}. However, studies on suspensions settling in 
viscoplastic fluids under quiescent conditions are very sparse \citep{Khabazi2016} so whether this is a viable
approach is not clear.

In this rapid communication we present direct numerical simulations of non-colloidal particle suspensions, in the dilute limit, settling in quiescent
viscoplastic fluids. Building on the out of plane investigation of \citep{Frigaard2017a}, we investigate the
stability criterion as a function of solid volume fraction for the in plane flow and comment on the transition to
settling---which, as far as we are aware, has not been investigated previously.

\section{Mathematical formulation and solution}
\label{sec:numerics}

In contrast to Newtonian fluids, very little numerical work has been done on suspension sedimentation in
viscoplastic fluids.
Many successful strategies for large scale suspension simulation in Newtonian 
fluids~\citep{Brady1988} rely on superposition
principles to make large scale computations tractable, which are not applicable to the non-linear viscoplastic
system. Coarse-grain approaches for Newtonian fluids use lubrication force models as sub-grid-scale models for the
under-resolved particle interactions. However, the authors recently found that such lubrication models 
cannot be straightforwardly applied in cases where particles are strongly confined in their yield envelopes such
as may occur in sedimentation without imposed shear \citep{Koblitz2018}.
This necessitates direct numerical simulation.

We consider the quasi-steady approximation of inertia-less, rigid circular particles suspended in 
incompressible viscoplastic fluid.
The fluid has velocity $\hat{\uv}(\hat{\xv})$, pressure $\hat{p}(\hat{\xv})$, plastic viscosity $\hat{\eta}$, 
and a total stress tensor
$\hat{\tauv}-\hat{p}\boldsymbol{\delta}$, where variables with a hat are dimensional.
In the absence of both fluid and particle inertia, and taking the particle buoyancy, $(\hat{\rho}_p -
\hat{\rho}_f)\hat{g}\hat{l}'$, as a characteristic stress scale $\mathcal{T}$--- where $\hat{\rho}_p$ 
and $\hat{\rho}_f$ are the particle and fluid densities, respectively, 
and $\hat{l}'$ is a characteristic length scale governed by the particle volume
to frontal area ratio \citep{Tokpavi2009}--we solve the non-dimensional steady Stokes equations
\begin{equation}
    \boldsymbol{\nabla}\cdot\boldsymbol{\tau}
    - \boldsymbol{\nabla} p = \dfrac{\rho_r}{1-\rho_r}, \quad
    \boldsymbol{\nabla}\cdot\boldsymbol{u} = 0, \label{EQ: mom dimensionless}
\end{equation}
where $\rho_r=\hat{\rho}_f/\hat{\rho}_p$ and the particles are negatively buoyant, such that $\rho_r<1$. 
We impose no-slip and no-penetration boundary conditions on the domain walls and particle surfaces. 
Additionally, we satisfy zero force and torque constraints, $\boldsymbol{F} = 0$, $\boldsymbol{T} = 0$, 
on the particles by adjusting their translational and rotational velocities. 
Here, $\boldsymbol{F}$ and $\boldsymbol{T}$ are defined for any given particle by

\begin{equation}
    \boldsymbol{F} =
    \int_\Gamma(-p\boldsymbol{n}+\boldsymbol{\tau}\cdot\boldsymbol{n})\,\mathrm{d}s+\boldsymbol{f}_b,\quad
    \boldsymbol{T} = 
    \int_\Gamma(\boldsymbol{x}-\boldsymbol{x}_b)\times(-p\boldsymbol{n}+\boldsymbol{\tau}\cdot
    \boldsymbol{n})\,\mathrm{d}s+\boldsymbol{t}_b,
\end{equation}
where $\Gamma$ denotes the particle surface, $\boldsymbol{x}$ a point on $\Gamma$, $\boldsymbol{x}_b$ the
position of the centre of mass, $\boldsymbol{n}$ the unit normal vector to the body surface, and 
$\boldsymbol{f}_b$ and $\boldsymbol{t}_b$ are external body force and torque, respectively.

We close the system using the ideal Bingham constitutive law

\begin{equation}
    \begin{cases}
        \boldsymbol{\tau}= \left(2 + \dfrac{Y}{||\dot{\gammav}||}\right)\dot{\boldsymbol{\gamma}} &
        \quad\mbox{if $||\tauv||>Y$,} \\
        \dot{\boldsymbol{\gammav}} = 0 & 
    \quad\mbox{if $\,||\tauv||\leq Y$,}
    \end{cases}
    \label{EQ: constitutive equation dimensionless}
\end{equation}

where the shear rate is defined as $\dot{\boldsymbol{\gamma}} := \frac{1}{2}\left(\nabla\boldsymbol{u}
    + \nabla\boldsymbol{u}^\intercal\right)$, and $||\cdot||$ are the induced norms of 
    the Frobenius inner product, defined as $||\mathsf{A}|| := \sqrt{1/2\mathsf{A}\cddot\mathsf{A}}$.
The force-free and torque-free conditions imply
that the particles adjust their velocities and angular velocities instantaneously \citep{Feng1995}. 

Unless otherwise stated, we use $\mathcal{T}/\hat{\eta}\hat{\rho}_f$,
$\mathcal{T}\hat{l}'/\hat{\eta}\hat{\rho}_f$, and
$\hat{R}$ as characteristic strain $\mathcal{G}$, velocity $\mathcal{V}$, and length 
$\mathcal{L}$ scales, respectively, where $\hat{R}$ is the particle radius and $\hat{l}'=\pi\hat{R}/2$.

Lastly, we define the critical yield number at which motion for a suspension of volume fraction $\phi$ stops as $Y^\ast_\phi$,
where for a single particle we find $Y^\ast_0=0.084$, in agreement with the literature
\citep{Tokpavi2008,Randolph1984,Chaparian2018} (note that \citet{Chaparian2018} use a different length scale in
their definition of $Y$).

We solve \eqref{EQ: mom dimensionless}--\eqref{EQ: constitutive equation dimensionless} using the 
widely adopted alternating direction multiplier method---also known as ALG2---developed by \citet{Glowinski1984}. 
It is extensively used in the literature, see \citet{Yu2007,Chaparian2017b,Muravleva2015} 
and references therein, so we do not give details here. We follow the implementation of \citet{Olshanskii2009,Muravleva2008}.

\citet{Wachs2016} studied the problem
numerically for a single sedimenting particle in a Bingham fluid, with a particular focus on the critical yield
stress required for cessation of motion. Even with a single particle in two dimensions with a relatively small
mesh (61440 cells), their computations took around 12 hours (wall-time). 
The limiting problem is the high computational expense of the ALG2 algorithm typically used for viscoplastic flow problems; 
discussion of ALG2 and related algorithms can be found in~\cite{Glowinski2014}. 
In order to make the larger problem sizes considered in this study tractable we adopt a quasi-steady
approximation, as in \citep{Chaparian2018}, and employ an overset grid discretization strategy. The grid
generation procedure is discussed at length in~\citep{Chesshire1990,Henshaw1998}, and has been used for
Newtonian and viscoplastic particulate flow problems by the authors in~\citep{Koblitz2017,Koblitz2018}. The
resultant linear system is inverted using the MUMPS massively parallel direct linear solver library \citep{Amestoy2001}.
Recently some significantly faster algorithms have been
developed~\citep{Saramito2016,Treskatis2016,Bleyer2018,Dimakopoulos2018}, and so we expect a rapid 
expansion in the near future of the size of problem which can be attempted.

\section{Results and discussion}

We compute the instantaneous velocity field for pseudo-random configurations of infinite circular cylinders, of
diameter $D$, in a confined domain with dimensions $(20\,D,120\,D)$, for solid volume fractions $\phi=(0.01,
0.05)$ over a yield number range of $Y=(0,0.173)$. Five realizations are computed at
each yield number, for each volume fraction. 

\begin{figure}
    \centering
    \includegraphics[scale=.9]{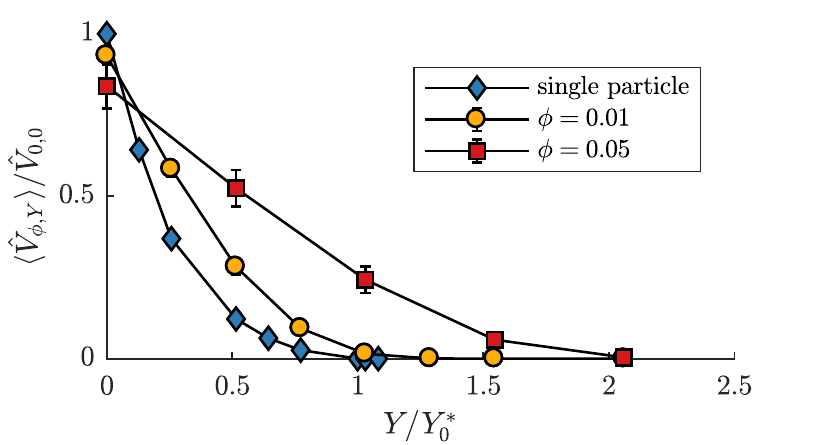}
    \caption{Mean sedimentation velocity at increasing yield numbers for $\phi=(0,0.01,0.05)$.\label{FIG:
    settling velocity}}
\end{figure}

\begin{figure}
    \centering
    \begin{minipage}[t]{.49\textwidth}
        \includegraphics[scale=.9]{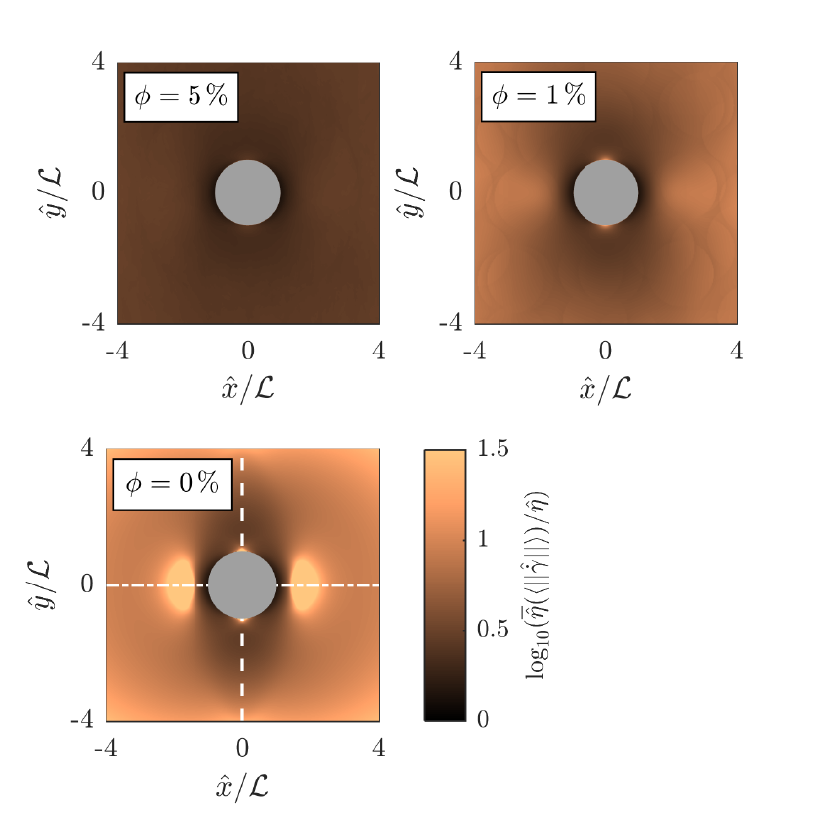}
    \end{minipage}%
    \hfill
    \begin{minipage}[t]{.49\textwidth}
        \includegraphics[scale=.9]{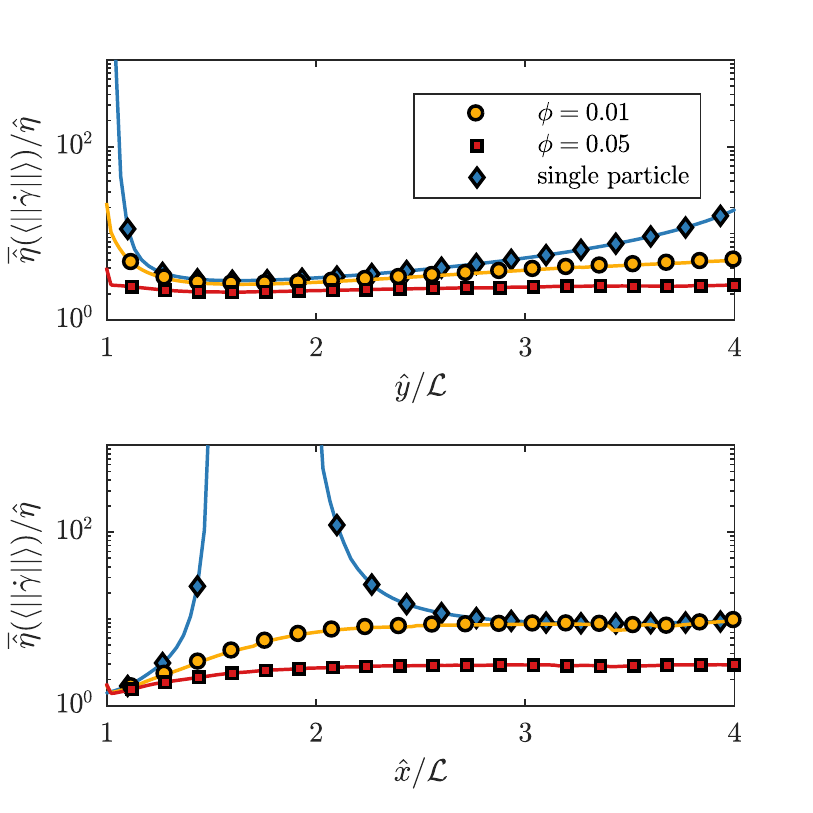}
    \end{minipage}%
    \caption{ 
    Left: Average effective
    viscosity near the particle surface for $Y=0.043$, with longitudinal (dashed line) and lateral (dash-dot line) axes of symmetries indicated in the
    $\phi=0$ panel. Right: average effective viscosity along the longitudinal (top) and lateral (bottom) axes 
    of symmetry for $Y=0.043$.
    \label{FIG: effective viscosity}}
\end{figure}

Figure~\ref{FIG: settling velocity} shows the mean settling velocity
of the suspension $\langle\hat{V}_{\phi,Y}\rangle$---normalized by the Stokes
velocity of a single particle in a Newtonian fluid $\hat{V}_{0,0}$---averaged over all
realizations, for increasing yield number. Here, the yield number is normalized using the critical yield number
required to hold a single particle at rest, $Y^\ast_0$.

All volume fractions show a decrease in settling velocity as the yield number increases. This is expected from
equation~\ref{EQ: constitutive equation dimensionless} where we can see that for any finite strain rate, a non-zero
yield number
leads to an increase in the effective viscosity. Looking at the limiting high yield number behaviour, we can see that
the critical yield number required to hold the suspension at rest, $Y^\ast_\phi$ increases with the solid volume fraction, 
as was found in \citet{Frigaard2017a} for out-of-plane settling. We find an almost two-fold increase in the
critical yield number
for $\phi=0.05$.

As $Y\rightarrow 0$ the medium mirrors a Newtonian fluid
with viscosity $\hat{\eta}$, showing hindered settling with increased $\phi$ \citep{Richardson1954}.
However, for $Y$ sufficiently large, we find higher settling efficiency with increasing solid volume
fraction, indicated by the increased mean settling velocity. This is similar to what has been observed in 
experimental studies of settling suspensions in
shear-thinning fluids \citep{Moreira2017}. 
There, settling particles shear fluid, causing a local decrease in viscosity and thereby
allowing nearby particles to settle more easily. 

We investigate possible shear thinning by examining the average local viscosity in the vicinity of a particle at a given
volume fraction at the same yield number, $Y=0.043$, where all particles for all volume fractions
are mobile. A rectangular grid is placed around each particle in the suspension
on which $||\hat{\dot{\gammav}}||$ is found by means of bilinear interpolation. This is then averaged over all particles in the
suspension, and used to find the average local effective viscosity,
$\overline{\hat{\eta}}(\langle||\hat{\dot{\gammav}}||\rangle)$, where $\langle\cdot\rangle$ denotes an average
over all particles. This is plotted as 
heat maps in the left panel of figure~\ref{FIG: effective viscosity}. The viscosity 
field for the single particle is as expected: arbitrarily high viscosity peaks (truncated in the plot) at the
unyielded equatorial plugs and the unyielded end-caps. As the volume fraction increases the viscosity field
becomes more uniform and is found to decrease in magnitude. The viscosity along the two axes of symmetry shown in
the $\phi=0$ plot of figure~\ref{FIG: effective viscosity} is further examined in the 
right panel of figure~\ref{FIG: effective viscosity}. The average viscosity decrease with 
increased solids volume fraction is evident,
and the viscosity field is found to be, on average, uniform $\mathcal{L}$ away 
from the particle surface. Near the surface, the
viscosity is high in the longitudinal direction due to the attached unyielded end-caps; however, in the lateral
direction the viscosity is low near the surface due to the viscoplastic boundary layer.

\begin{figure}
    \centering
    \begin{minipage}[t]{.49\textwidth}
        \includegraphics[width=\textwidth]{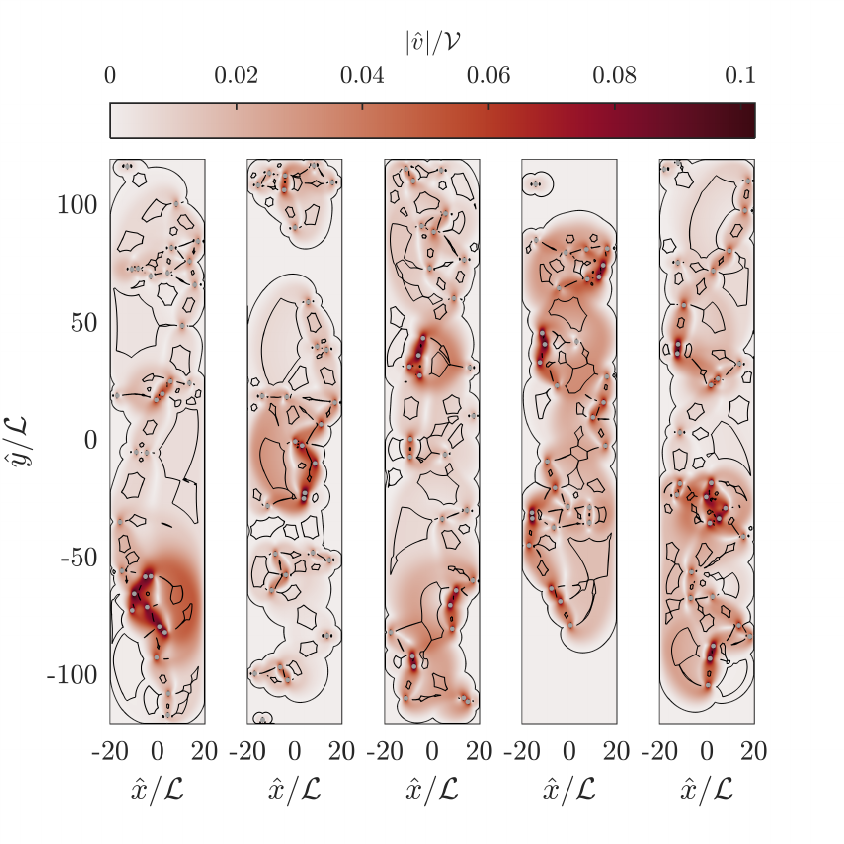}
    \end{minipage}%
    \begin{minipage}[t]{.49\textwidth}
        \includegraphics[width=\textwidth]{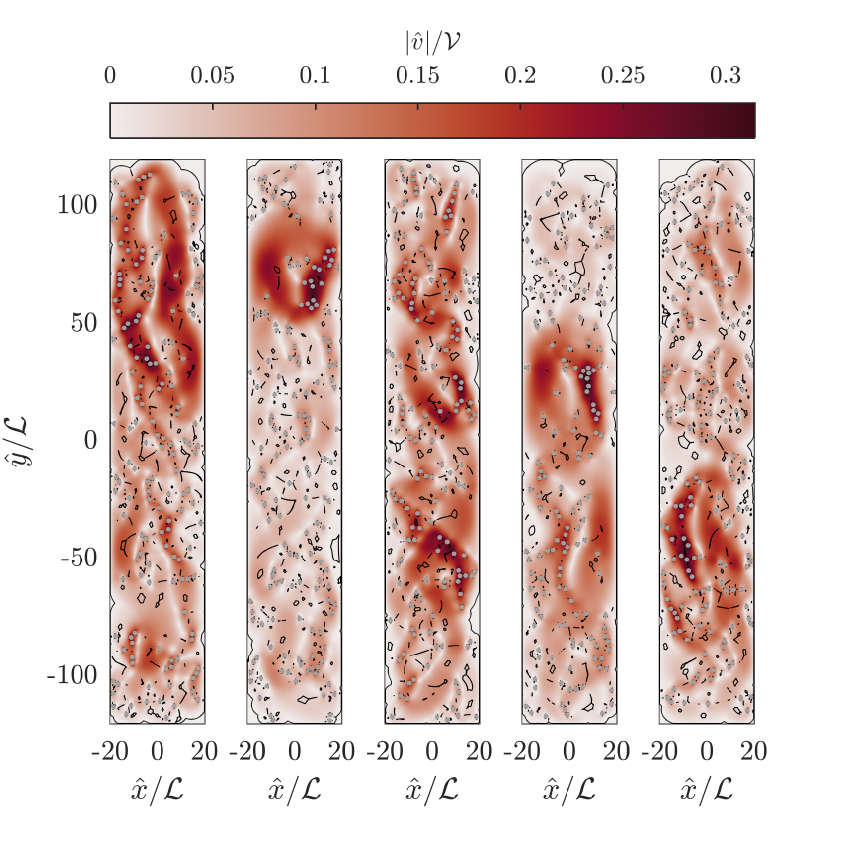}
    \end{minipage}%
    \caption{Heat maps of the fluid velocity magnitude with particles overlaid in grey and yield surface indicated
    by the solid black lines for $Y=0.087$ with $\phi=0.01$ (left) and $\phi=0.05$ (right) for five different realizations.
    \label{FIG: velocity figures}}
\end{figure}

\begin{figure}
    \centering
    \begin{minipage}[t]{.49\textwidth}
    \includegraphics[width=\textwidth]{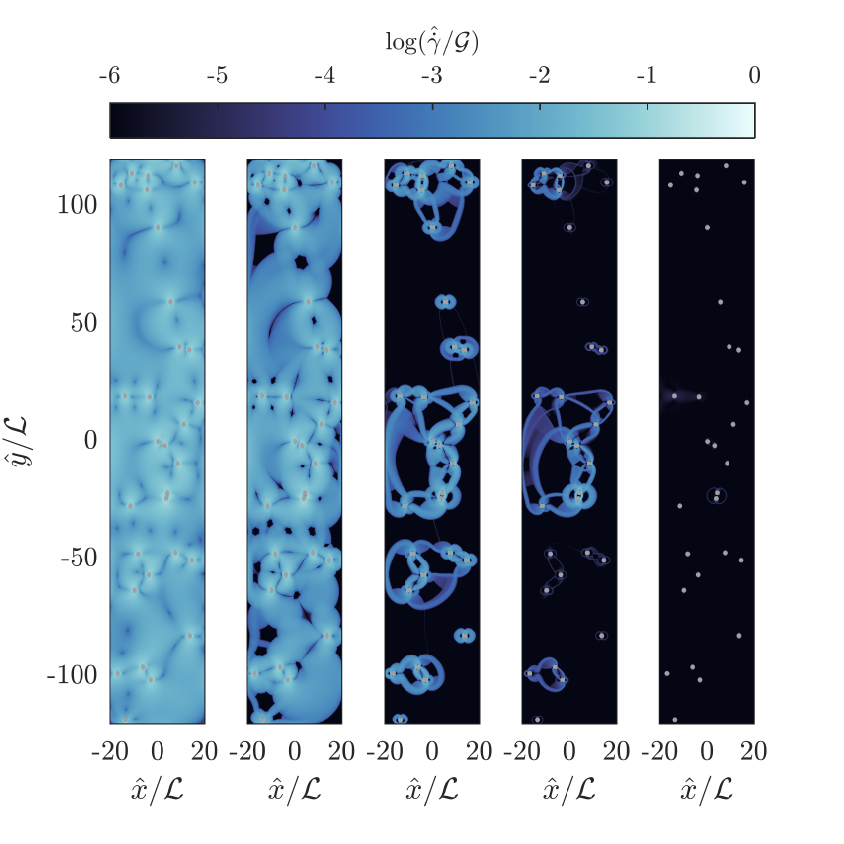}
    \end{minipage}%
    \hfill
    \begin{minipage}[t]{.49\textwidth}
    \includegraphics[width=\textwidth]{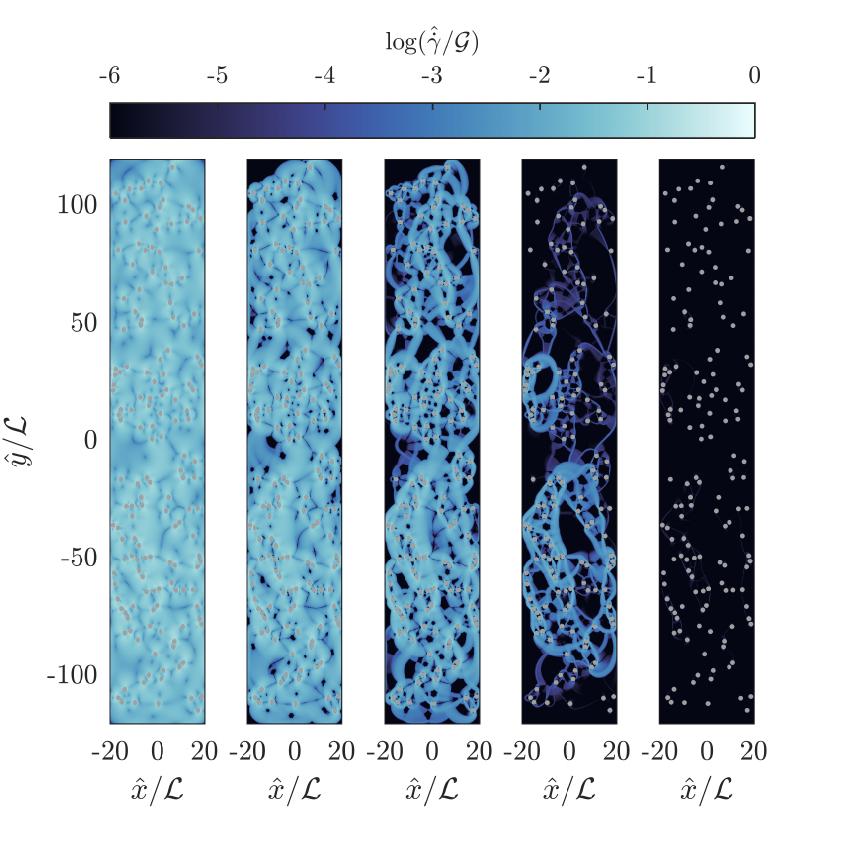}
    \end{minipage}
        \caption{Heat maps of $\log(\dot{\gamma}/\mathcal{G})$ for $\phi=0.01$ at $Y=(0, 0.022,
        0.065, 0.087, 0.13)$ (left panel) and $\phi=0.05$ at $Y=(0, 0.043, 0.087, 0.13, 0.17)$ (right panel). \label{FIG: deformation}}
\end{figure}

It has been shown experimentally that static suspensions in yield stress fluids will settle 
when a macroscopic stress, for example by shearing the system, is introduced~\citep{Ovarlez2012,Merkak2009}. 
In the absence of a macroscopic flow, only the buoyancy stress of the individual particles acts on the fluid to
drive the flow. 
Studies of small systems of particles in the resistance formulation demonstrated a drag reduction 
for inline particle arrangements \citep{Liu2003,Yu2007,Tokpavi2009}. Equivalently, the recent study 
of~\citet{Chaparian2018} demonstrated higher
velocities for inline configurations in the mobility formulation. The current simulations corroborate this in so
far as structures of vertically clustered particles can clearly be visually identified to be settling fastest, for
example, see the velocity magnitude heat maps in figure~\ref{FIG: velocity figures}.
By examining the velocity field \citet{Chaparian2018} demonstrated that nearby particles had little effect on
one another beyond a relatively small separation distance $\hat{d}_{\text{sep}}/\mathcal{L}<20$, suggesting that the
stress decay in a viscoplastic fluid is appreciably faster than in a Newtonian fluid where $||\tauv||\sim r^{-1}$ as
$r\rightarrow\infty$ \citep{Tanner1993}.
As the yield number
increases the rapid stress decay becomes more relevant. In the left panel of figure~\ref{FIG: deformation} 
the logarithmically-scaled heat map of the strain rate magnitude is shown for a realization of $\phi=0.01$, 
with the yield number increasing
from left to right. As expected, unyielded material emerges as the yield number increases. For
$Y\gtrsim0.065$ discrete yield envelopes surround particles, coalescing into larger pockets
when particles are close together.
As the yield number increases further we find clusters of particles inside yield envelopes with
comparatively isolated particles held static. For $\phi=0.05$ we see a similar trend in the 
$\log(\dot{\gammav})$ heat maps shown in the right panel of figure~\ref{FIG: deformation}. However, 
there are fewer distinct groups
of particles in individual yield envelopes. It is likely that the computational domain is not sufficiently large
for a suspension this dense. 

We take a closer look at the suspension morphology near $Y^\ast_0$ in
figure~\ref{FIG: velocity figures}, where heat maps of the velocity magnitude with yield surface overlays are
shown for all five realizations for $\phi=0.01$ (left panel) and $\phi=0.05$ (right panel) at $Y=0.087$. 
All realizations for $\phi=0.01$
show both static particles and yield pockets with sedimentation clusters, while for $\phi=0.05$ no static
particles are present in any realization and velocity peaks are found near vertically arranged clusters.

\begin{figure}
    \centering
    \includegraphics[scale=.9]{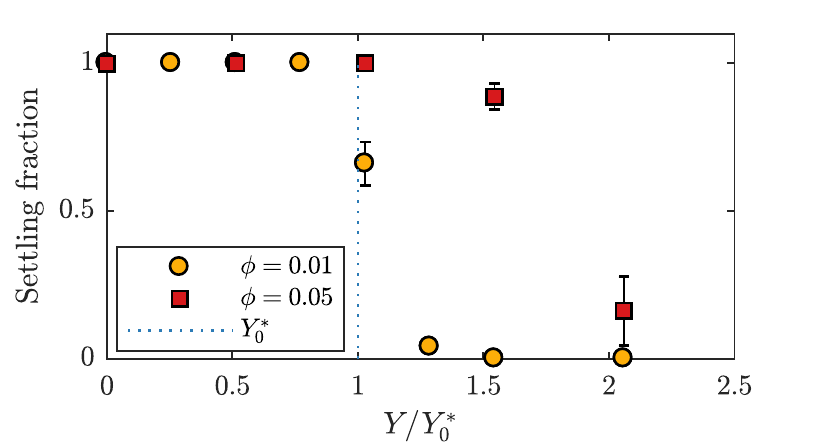}
    \caption{Proportion of settling
    particles in the suspension at increasing yield strengths with the critical yield number for a single
    particle indicated by the dashed line.\label{FIG: settling fraction}}
\end{figure}

This formation of settling pockets of particle-dense regions and isolated
static particles is not evident from the mean suspension velocity. In figure~\ref{FIG: settling fraction} we show the proportion of
settling particles for both volume fractions as the yield number increases, averaged over all
realizations. In both cases, for $Y/Y^\ast_0<1$ the entire suspension settles as the yield number is below the
threshold required to hold even a single particle static. At $Y/Y^\ast_0\approx1$ the vast majority of the
suspension settles for both cases, with a fraction of static particles evident for $\phi=0.01$. As the yield
number increases beyond $Y/Y^\ast_0=1$ more isolated particles are held static, leading to a decrease in the settling
fraction for both volume fractions considered. For $\phi=0.01$ we find that the entire suspension is held static
at $Y/Y^\ast_0\approx1.5$. The critical yield number required to
hold the entire $\phi=0.05$ suspension static was not reached in this study due to convergence time requirements.

Concentrating on $\phi=0.01$, for which we have found the critical yield number, we can see three distinct flow regimes: Regime (I)
for $0\leq Y<Y^\ast_0$ where the entire suspension is settling; Regime (II) for $Y^\ast_0\leq Y<Y^\ast_{\phi}$ where the
proportion of static particles increases as $Y\rightarrow Y^\ast_{\phi}$; Regime (III) for $Y\geq Y^\ast_{\phi}$, for
which the entire suspension is held stationary. While $Y^\ast_\phi$ was not reached for $\phi=0.05$ in this study we
have no reason to believe that the suspension will not be held static with a sufficiently high yield number.
Experimental studies have worked with statically held suspensions as dense as $\phi = 0.4$~\citep{Ovarlez2012}.

\begin{figure}
    \centering
    \begin{minipage}[t]{.49\textwidth}
    \includegraphics[scale=1.]{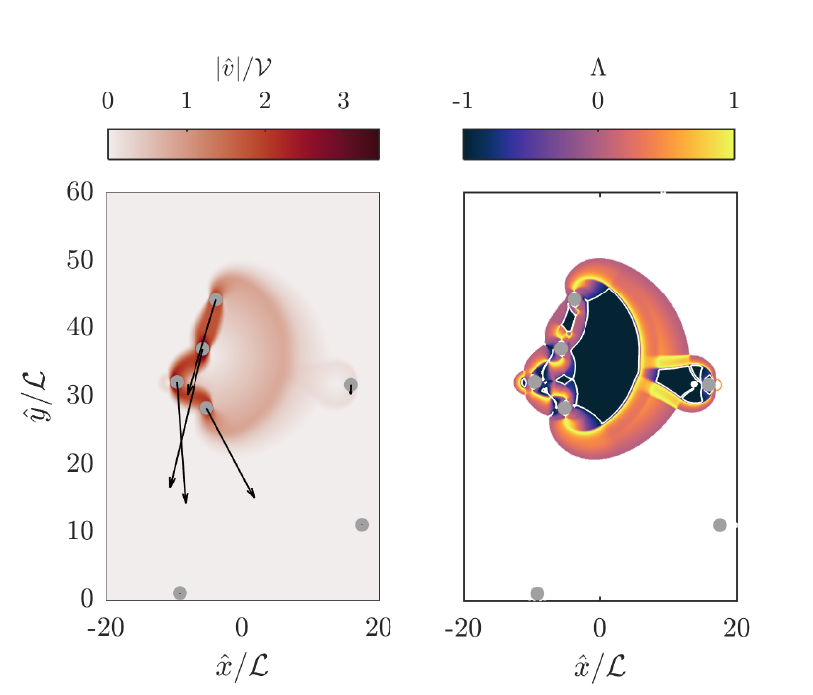}
    \end{minipage}%
    \hfill
    \begin{minipage}[t]{.49\textwidth}
    \includegraphics[scale=1.]{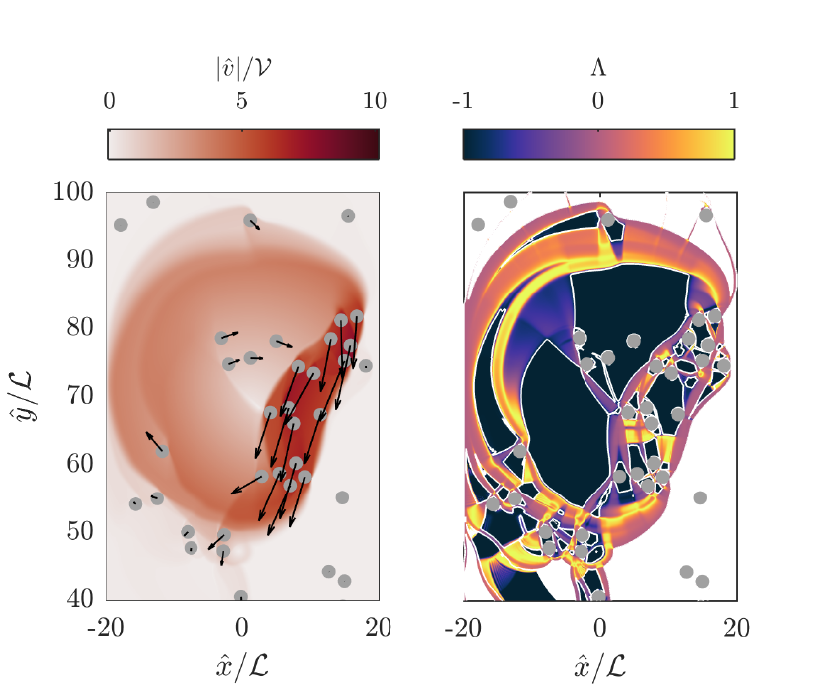}
    \end{minipage}
    \caption{Examples of the complex flow features found in regime (II). Left: Close-up view from a $\phi=0.01$
    suspension with $Y=0.087$ showing the mobilization of an isolated particle
    by a passing sedimentation cluster. Right: Close-up view from a $\phi=0.05$ suspension with $Y=0.173$
    showing lone particles swept up in the strong recirculating flow generated
    by a sedimentation cluster. A large unyielded region undergoing rigid body motion with embedded particles is
    visible. For each panel, the left plots show heat maps of the fluid velocity magnitude, with particles
    overlaid in grey and their individual velocity vectors indicated by arrows. The right plots show heat maps of
    the normalized second invariant of the velocity gradient tensor, with static unyielded regions masked off,
    yield surfaces indicated by the solid white lines, and particles overlaid in grey. \label{FIG: zoom figs}}
\end{figure}

In regime (I) the fluid could be considered as weakly
shear-thinning and regime (III) is trivial, while in regime (II) the strong competition between the yield stress
and buoyancy of groups of particles leads to complex flow features, which the larger error bars above
$Y/Y^\ast_0>1$ in figure~\ref{FIG: settling fraction} indicate. Two examples of the complex flow features found in this
regime are shown in figure~\ref{FIG: zoom figs}. For each of the two features the left plots show heat maps of
the fluid velocity magnitude, 
while the right plots show heat maps of the normalized second invariant of the velocity gradient tensor,
$\Lambda$---a metric used to describe the character of the flow. Negative and positive values of $\Lambda$ show
where flow is dominated by enstrophy and strain, where for $\Lambda=-1$ flow is purely rotational, $\Lambda=0$
flow undergoes simple shear, and $\Lambda=+1$ flow is purely extensional \citep{Hemingway2018}. 
The left panel shows a sedimentation cluster in a $\phi=0.01$ suspension mobilizing a lone 
particle as it moves past. The right panel is from a $\phi=0.05$ suspension showing a collection of particles
settling together, generating a strong recirculating flow within the yield envelope. Lone particles are swept up
by this recirculating flow, and two large unyielded plugs can be seen, one of which has embedded particles
that would have otherwise been held static. It is flow features such as these that lead to the larger error bars
in figure~\ref{FIG: settling fraction} but also complicate the identification of spatial correlation lengths, for
example, critical separation distances between particles that encourage settling at yield numbers beyond
$Y^\ast_0$.

\section{Conclusions}

In this rapid communication we investigated settling of two-dimensional non-colloidal particles in viscoplastic fluids under
quiescent conditions by means of direct numerical simulation.
Three flow regimes were
identified, where (I) the entire suspension settles, (II) there exist both static and settling particles in the
same suspension, and (III) the entire suspension is arrested.

In regime (I) for sufficiently high yield numbers we observe enhanced settling with increased volume fraction,
opposite to a Newtonian fluid, which we attribute to shear-thinning. Regime (II) displays complex flow features
such as sedimenting clusters, mobilization of lone particles, and rigid recirculating zones. For suspension volume
fractions greater than zero the transition to regime (III) is delayed, requiring a higher yield number to hold the
suspension static than is required to hold a single particle static. This corroborates the 
theoretical work of \citet{Frigaard2017a} and the inferences drawn from
studies of small-scale model systems \citep{Tokpavi2009,Chaparian2018}.

Further research is required to explore the dynamics of the settling phases in regimes (I) and (II). Regime (II)
is of particular interest since it may lead to heterogeneities in the suspension. 
Understanding this regime and the transition to regime (III) may play a role in explaining observations in many
industrial and natural problems involving the sedimentation of viscoplastic suspensions under quiescent flow
conditions.
We anticipate that new solution methods \citep{Treskatis2016,Bleyer2018,Saramito2016,Dimakopoulos2018} will soon enable much larger simulations of
such systems, including the ability to investigate time-evolution and inertia, which were neglected in this study.
Finally, we encourage researchers to investigate this problem experimentally.

\bibliography{suspension_paper}

\end{document}